\documentclass[12pt]{iopart}
\usepackage{iopams}
\usepackage{amssymb}
\usepackage{epsfig}

\newcommand{\mb}[1]{\mbox{\boldmath $#1$}}
\begin{document}

\title[Rotating Collapse of Stellar Iron Cores in General Relativity]
{Rotating Collapse of Stellar Iron Cores in \\ General Relativity}

\author{C D Ott$^{1,2}$,
  H Dimmelmeier$^3$, A Marek$^3$, H-T Janka$^3$, B Zink$^4$,
  I Hawke$^5$ and E Schnetter$^4$}
\ead{cott@as.arizona.edu}

\address{$^1$ Max-Planck-Institut f\"ur Gravitationsphysik,
  Albert-Einstein-Institut, \\
  Am M\"uhlenberg~1, 14476 Potsdam, Germany}

\address{$^2$ JINA Postdoctoral Fellow, 
  Department of Astronomy and Steward Observatory, \\
  The University of Arizona, Tucson, AZ 85721, USA}

\address{$^3$ Max-Planck-Institut f\"ur Astrophysik, \\
  Karl-Schwarzschild-Strasse~1, 85741 Garching, Germany}

\address{$^4$  Center for Computation \& Technology, Louisiana State University, \\
  216 Johnston Hall, Baton Rouge, LA 70803, USA}

\address{$^5$ School of Mathematics, University of Southampton,
  Southampton SO17 1BJ, UK}

%%%%%%%%%%%%%%%%%%%%%%%%%%%%%%%%%%%%%%%%%%%%%%%%%%%%%%%%%%%%%%%%%%%%%%%%%%%%%%%%%%%
%%%%%%%%%%%%%%%%%%%%%%%%%%%%%%%%%%%%%%%%%%%%%%%%%%%%%%%%%%%%%%%%%%%%%%%%%%%%%%%%%%%
%%%%%%%%%%%%%%%%%%%%%%%%%%%%%%%%%%%%%%%%%%%%%%%%%%%%%%%%%%%%%%%%%%%%%%%%%%%%%%%%%%%

\begin{abstract}
  We present results from the first $2+1$ and $3+1$ simulations of the
  collapse of rotating stellar iron cores in general relativity
  employing a finite-temperature equation of state and an approximate
  treatment of deleptonization during collapse. We compare full
  $3+1$ and conformally-flat spacetime evolution methods and find
  that the conformally-flat treatment is sufficiently accurate for the
  core-collapse supernova problem. We focus on the gravitational wave
  (GW) emission from rotating collapse, core bounce, and early
  postbounce phases. Our results indicate that the GW signature of
  these phases is much more generic than previously estimated. In
  addition, we track the growth of a nonaxisymmetric instability of
  dominant $ m = 1 $ character in two of our models that leads to
  prolonged narrow-band GW emission at
  {$ \sim 920 \mbox{\,--\,}930 \mathrm{\ Hz} $} over several tens of
  milliseconds.
\end{abstract}


%%%%%%%%%%%%%%%%%%%%%%%%%%%%%%%%%%%%%%%%%%%%%%%%%%%%%%%%%%%%%%%%
%%%%%%%%%%%%%%%%%%%%%%%%%%%%%%%%%%%%%%%%%%%%%%%%%%%%%%%%%%%%%%%%
%%%%%%%%%%%%%%%%%%%%%%%%%%%%%%%%%%%%%%%%%%%%%%%%%%%%%%%%%%%%%%%%

\section{Introduction}
\label{section:intro}

For more than two decades astrophysicists have struggled to compute
the gravitational wave (GW) signal produced by rotating stellar iron core
collapse and the subsequent supernova evolution. Besides the
coalescence of black hole and neutron star binaries, core-collapse
events are considered to be among the most promising sources of
detectable GWs. Theoretical predictions of the core-collapse
supernova GW signature are still hampered by three
major problems: (i) The rotational configuration prior to
gravitational collapse is still uncertain since multi-D evolutionary
calculations of rotating massive stars have not yet been performed;
(ii) reliable waveform estimates require a general relativistic (GR)
treatment, since both high densities and high velocities in combination
with strong gravitational fields are encountered in this problem; and
(iii) an adequate treatment of the nuclear equation of state (EOS) and
the neutrino microphysics/radiative transfer is crucial for obtaining
realistic collapse, bounce, and postbounce dynamics and waveforms. GW
emission from core-collapse supernovae may arise from rotating
collapse and bounce, postbounce neutrino-driven convection,
anisotropic neutrino emission, nonaxisymmetric rotational
instabilities of the protoneutron star (PNS), or from the recently
proposed PNS core g-mode oscillations~\cite{burrows_06_a}. 
Previous estimates of the GW signature of core-collapse 
supernovae have relied either on Newtonian
simulations~\cite{moenchmeyer_91_a, zwerger_97_a, kotake_03_a,
  ott_04_a, ott_06_b}
(to some extent approximating GR effects~\cite{mueller_04_a,
obergaulinger_06_a}),
or GR simulations with simplified analytic (so-called hybrid) EOSs and
no neutrino treatment~\cite{dimmelmeier_02_b, shibata_04_a,
  shibata_05_a, cerda_05_a}.
Depending on the rotation strength, the softness of the EOS at subnuclear
densities, and the inclusion of GR effects, the collapse dynamics and,
accordingly, the GW signatures can differ significantly.

In previous studies, at least three ``types'' of dynamics and
resulting GW signatures of rotating collapse were identified:
Type~I dynamics is characterized by little influence of
centrifugal effects during collapse and bounce. Type~I
models undergo core bounce governed by the stiffening
of the nuclear EOS at nuclear density and ``ring down'' 
quickly into postbounce equilibrium. Their waveforms
exhibit one pronounced large (negative) spike at bounce 
and then show a gradually damped ring-down wave signal at
early postbounce times. Type~II models, on the
other hand, are rotation dominated and undergo core bounce
at densities below (or slightly above) nuclear matter density under
the strong influence of centrifugal forces. Their dynamics
generally exhibits multiple harmonic-oscillator-like damped slow
cycles of coherent bounce--re-expansion--collapse (``multiple bounces''),
which is reflected in the waveform by distinct signal peaks associated
with every bounce. Type~III dynamics is characterized by 
fast collapse (owing to very efficient electron capture for instance
in accretion-induced collapse or large artificial initial pressure
reduction in polytropic core-collapse models), extremely small
masses of the homologously collapsing inner core, and low-amplitude GW emission.

In this article we present new results from GR simulations, 
focussing on the rotating collapse, bounce, and early postbounce 
phases. As a first stage we perform 2D (axisymmetric)
and 3D calculations with polytropic initial models and a 
hybrid EOS in order to compare results obtained in the
conformally-flat approximation to GR (CFC; conformal-flatness
condition~\cite{isenberg_78_a, wilson_96_a}) with results
from full Cauchy free evolutions in $3+1$ GR. The second stage
encompasses the first ever multi-D simulations of rotating
stellar iron core collapse in GR with presupernova models
from stellar evolution calculations, a finite-temperature nuclear EOS,
and a simple, but effective treatment of electron capture and neutrino
radiation effects during collapse. In this way, we obtain the  
most accurate estimates of the GW signature of rotating stellar core
collapse in full GR to date. 

This article is organized as follows. In \sref{section:methods},
we delineate the methods and numerical tools employed and present
the initial model data that we utilize in our calculations.
In \sref{section:results}, we present the numerical results of
our calculations, focussing on the CFC--full-GR comparison
in \sref{section:hybrid_comparison}, on the GW signature of axisymmetric rotating
core collapse with microphysics 
in \sref{section:shen_comparison_axi}, and on the development of 
nonaxisymmetric structures and dynamics in \sref{section:shen_comparison_nonaxi}.
In \sref{section:summary}, we summarize our results
and discuss the detectability of the GWs 
from our models.

%%%%%%%%%%%%%%%%%%%%%%%%%%%%%%%%%%%%%%%%%%%%%%%%%%%%%%%%%%%%%%%%%%%%%%%%%%%
%%%%%%%%%%%%%%%%%%%%%%%%%%%%%%%%%%%%%%%%%%%%%%%%%%%%%%%%%%%%%%%%%%%%%%%%%%%
%%%%%%%%%%%%%%%%%%%%%%%%%%%%%%%%%%%%%%%%%%%%%%%%%%%%%%%%%%%%%%%%%%%%%%%%%%%

\section{Methods and Initial Model Data}
\label{section:methods}

We base this work on
two independent numerical codes: the \textsc{CoCoNuT} code
of Dimmelmeier~et~al.~\cite{dimmelmeier_02_a, dimmelmeier_02_b, dimmelmeier_05_a}
which treats GR in the conformally-flat approximation and
which we employ in 2D (axisymmetric) mode and 
\textsc{Cactus/Carpet/Whisky} (CCW)~\cite{goodale_03_a, baiotti_04_a,
  ott_06_c, ott_06_d} for $3+1$ full free Cauchy-evolution GR
calculations. CCW uses mesh-refined Cartesian grids while
\textsc{CoCoNuT} operates in spherical coordinates. Both codes employ
the same formulation of GR hydrodynamics which is outlined
in the following \sref{section:gr_hydro}. Code details and the different
methods for time-updating the GR curvature fields are discussed
in \sref{section:ccw} for CCW and 
in \sref{section:coconut} for \textsc{CoCoNuT}. In 
\sref{section:eos}, we discuss the EOSs that we utilize, while
\sref{section:deleptonization} deals with the deleptonization
and neutrino pressure treatment that we implement in our codes.
In \sref{section:models} we introduce the initial model data
used for our calculations, and in \sref{section:wave_extraction} we
discuss the GW extraction method.

%%%%%%%%%%%%%%%%%%%%%%%%%%%%%%%%%%%%%%%%%%%%%%%%%%%%%%%%%%%%%%%%%%%%%%%%%%%
%%%%%%%%%%%%%%%%%%%%%%%%%%%%%%%%%%%%%%%%%%%%%%%%%%%%%%%%%%%%%%%%%%%%%%%%%%%

\subsection{GR Hydrodynamics}
\label{section:gr_hydro}

We adopt the ADM $ 3 + 1 $ foliation of spacetime~\cite{york_79_a}.
All equations assume $c = G = 1$. 
In the following, Latin indices run from 1 to 3 while Greek ones run
from 0 to 3. We adhere to abstract index notation. $g_{\mu\nu}$ is the
4-metric and $\gamma_{ij}$ is the 3-metric.

The hydrodynamic evolution of a perfect fluid in GR
with 4-velocity $ u^{\,\mu} $, rest-mass current
$ J^{\,\mu} = \rho u^{\,\mu} $, where $\rho$ is the rest-mass
density, and stress-energy tensor
$ T^{\mu \nu} = \rho h u^{\,\mu} u^{\,\nu} + P g^{\,\mu \nu} $ is
determined by a system of local conservation equations,
\begin{equation}
  \nabla_{\!\mu} J^\mu = 0, \qquad \nabla_{\!\mu} T^{\mu \nu} = 0,
  \label{eq:equations_of_motion_gr}
\end{equation}
where $ \nabla_{\!\mu} $ denotes the covariant derivative with respect
to the 4-metric. The quantity $ h = 1 + \epsilon + P / \rho $ 
is the specific enthalpy, $P$ is the fluid pressure, 
and the 3-velocity is given by 
$ v^{\,i} = u^{\,i} / (\alpha u^{\,0}) + \beta^i / \alpha $,
where $\alpha$ is the lapse function and $\beta^i$ is the coordinate 
shift. We define the set of conserved variables as
\begin{displaymath}
  D = \rho W,
  \qquad
  S^i = \rho h W^{\,2} v^i,
  \qquad
  \tau = \rho h W^{\,2} - P - D.
\end{displaymath}
In the above expressions $ W = \alpha u^0 $ is the Lorentz factor,
which satisfies the relation $ W = 1 / \sqrt{1 - v_i v^i} $.

The local conservation laws~(\ref{eq:equations_of_motion_gr})
are written as a first-order, flux-conservative system of
hyperbolic equations~\cite{banyuls_97_a},
\begin{equation}
  \frac{\partial \sqrt{\gamma\,} \mb{U}}{\partial t} +
  \frac{\partial \sqrt{- g\,} \mb{F}^{\,i}}{\partial x^{\,i}} =
  \sqrt{- g\,} \mb{S},
  \label{eq:conservation_equations_gr}
\end{equation}
with
\begin{eqnarray*}
  \mb{U} & = & [D, S_j, \tau], \\
  \mb{F}^{\,i} & = &
  \left[ D \hat{v}^{\,i}, S_j \hat{v}^{\,i} + \delta^{\,i}_j P,
  \tau \hat{v}^{\,i} + P v^{\,i} \right]\!, \\
  \mb{S} & = &
  \left[ 0, T^{\mu \nu} \left( \frac{\partial g_{\nu j}}{\partial x^{\,\mu}} - 
  \Gamma^{\,\lambda}_{\mu \nu} g_{\lambda j} \right)\!,
  \alpha \left( T^{\mu 0}
  \frac{\partial \ln \alpha}{\partial x^{\,\mu}} -
  T^{\mu \nu} \Gamma^{\,0}_{\mu \nu} \right) \right]\!.
\end{eqnarray*}%
Here $ \hat{v}^{\,i} = v^{\,i} - \beta^i / \alpha $, and $ g $ and
$ \gamma $ are the determinant of $ g_{\mu\nu} $ and $ \gamma_{ij} $,
respectively, with $ \sqrt{-g} = \alpha \sqrt{\gamma} $.
$ \Gamma^{\,\lambda}_{\mu \nu} $ are the 4-Christoffel symbols.

The above equations are solved in semi-discrete fashion. The spatial
discretization is performed by means of a high-resolution
shock-capturing (HRSC) scheme employing a second-order accurate
finite-volume discretization. We make use of the Marquina flux formula
for the local Riemann problems and piecewise-parabolic cell interface
reconstruction (PPM). For a review of such methods in the GR context,
see~\cite{font_03_a}. The time integration and coupling with curvature 
are carried out with the Method of Lines~\cite{hyman_76_a} in
combination with a second-order accurate explicit Runge--Kutta scheme
in \textsc{CoCoNuT} and a second-order accurate explicit iterated
Crank--Nicholson~\cite{teukolsky_00_a} integrator in CCW.

%%%%%%%%%%%%%%%%%%%%%%%%%%%%%%%%%%%%%%%%%%%%%%%%%%%%%%%%%%%%%%%%%%%%%%%%%%%
%%%%%%%%%%%%%%%%%%%%%%%%%%%%%%%%%%%%%%%%%%%%%%%%%%%%%%%%%%%%%%%%%%%%%%%%%%%

\subsection{\textsc{Cactus/Carpet/Whisky}}
\label{section:ccw}

In York's variant of the ADM $ 3 + 1 $ Cauchy-evolution
formalism~\cite{york_79_a}, the Einstein equations split into a
coupled set of first-order evolution equations for the 3-metric
$ \gamma_{ij} $ and the extrinsic curvature $ K_{ij} $,
\begin{eqnarray}
  \partial_t \gamma_{ij} & = & - 2 \alpha K_{ij} +
  \nabla_{\!i} \beta_j + \nabla_{\!j} \beta_i,
  \label{eq:adm_equations_1}
  \\
  \partial_t K_{ij} & = & - \nabla_{\!i} \nabla_{\!j} \alpha +
  \alpha \left( R_{ij} + KK_{ij} - 2 K_{ik} K_j^k \right) +
  \beta^k \nabla_{\!k} K_{ij}
  \label{eq:adm_equations_2}
  \\
  & & + K_{ik} \nabla_{\!j} \beta^k + K_{jk} \nabla_{\!i} \beta^k -
  8 \pi \alpha \left( S_{ij} - \frac{\gamma_{ij}}{2}
  \left( S_k^k - \rho_\mathrm{ADM} \right) \right),
  \nonumber
\end{eqnarray}%
and constraint equations,
\begin{eqnarray}
  0 & = & R + K^2 - K_{ij} K^{ij} - 16 \pi \rho_\mathrm{ADM},
  \label{eq:adm_equations_3}
  \\
  0 & = & \nabla_{\!i} \left( K^{ij} - \gamma^{ij} K \right) -
  8 \pi S^j.
  \label{eq:adm_equations_4}
\end{eqnarray}%

In the above equations,
$ R_{ij} $ is the 3-Ricci tensor and $ R $ is the
scalar 3-curvature. The projection of the stress-energy tensor onto
the spatial hypersurface is 
$ S_{ij} = \rho h W^2 v_i v_j + \gamma_{ij} P $,
the ADM energy density is given by $\rho_\mathrm{ADM} = \rho h W^2 - P $,
and $S^j = \rho h W^2 v^i$ is the momentum density as measured
by an Eulerian observer moving orthogonally to the spacelike 
hypersurfaces.

In CCW we employ the AEI-BSSN code that implements the BSSN recast
of the ADM equations in the fashion detailed in~\cite{alcubierre_99_a,
  alcubierre_03_a}. Here, we mention only that the BSSN formalism
makes use of a conformal decomposition of the 3-metric, 
 \hbox{$\tilde \gamma_{ij} = \mathrm{e}^{- 4 \phi}
\gamma_{ij}$}, and the trace-free part of the extrinsic curvature,
\hbox{$A_{ij} = K_{ij} - \gamma_{ij} K/3$}, with the conformal factor
$\phi$ chosen to satisfy $\mathrm{e}^{4 \phi} = \gamma^{1/3}$. In this
formulation, in addition to the evolution equations for the conformal
3-metric $\tilde \gamma_{ij}$ and the conformal traceless extrinsic
curvature $\tilde A_{ij}$, there are evolution equations for
the conformal factor $\phi$, for the trace of the extrinsic curvature $K$,
and for the ``conformal connection functions'' $\tilde \Gamma^i \equiv
\partial \tilde\gamma^{ij} / \partial x^j $. 

We employ the common $1+\log$ slicing condition and a $\Gamma$-driver
type condition  for the shift~\cite{alcubierre_03_a},
$\partial/\partial t\, \beta^i = (\tilde{\Gamma}^i + \Delta t\,
\partial/\partial t\, \tilde{\Gamma}^{i})$, which is similar
to that used in~\cite{shibata_04_a}.

Mesh refinement in CCW is provided by the \textsc{Carpet}
driver~\cite{schnetter_03_a} which implements
full Berger--Oliger mesh refinement~\cite{berger_84_a} with subcycling in time.
In the calculations presented here we employ box-in-box 
refinement hierarchies with a maximum of 9 levels
of refinement and a factor of 2 increase in resolution
from one level to the next. We run 
\textsc{Carpet} in progressive mesh refinement mode
and activate predefined refinement levels based on a 
density criterion as the collapse proceeds~\cite{ott_06_d}.
The grids extend to a maximum outer diagonal radius of 
$\sim 5000 \mathrm{\ km}$ and the finest grid has a linear
zone size of $\sim 350 \mathrm{\ m}$.

%%%%%%%%%%%%%%%%%%%%%%%%%%%%%%%%%%%%%%%%%%%%%%%%%%%%%%%%%%%%%%%%%%%%%%%%%%%
%%%%%%%%%%%%%%%%%%%%%%%%%%%%%%%%%%%%%%%%%%%%%%%%%%%%%%%%%%%%%%%%%%%%%%%%%%%

\subsection{\textsc{CoCoNuT}}
\label{section:coconut}

In the \textsc{CoCoNuT} code we employ the CFC approximation
to GR introduced by Isenberg~\cite{isenberg_78_a} and first 
used in a pseudo-evolutionary context by Wilson et
al.~\cite{wilson_96_a}. Details on the CFC GR equations and
implementation specifics can be found in \cite{dimmelmeier_02_a,
  dimmelmeier_05_a}. Here we mention only the salient features of CFC
in which the ADM spatial 3-metric is replaced by the conformally-flat
3-metric, $\gamma_{ij} = \phi^4 \hat{\gamma}_{ij}$, where
$\hat{\gamma}_{ij}$ is the flat-space metric. Hence
$\hat{\gamma}_{ij} = \delta_{ij}$ in Cartesian coordinates and
$\hat{\gamma}_{ij} = \mathrm{diag}\, (1,r^2, r^2 \sin\theta)$ in 
spherical coordinates. Here $\phi$ is the conformal factor.

In the CFC approximation the ADM
equations~(\ref{eq:adm_equations_1}--\ref{eq:adm_equations_4})
reduce to a set of elliptic equations for $ \phi $, $\alpha$, and
$\beta^i$, if additionally maximal slicing is assumed: $K=0$. The extrinsic
curvature becomes a function of these variables and the shift
is to leading order (for matter-dominated spacetimes) 
identical to the minimal distortion condition~\cite{york_79_a}.
The CFC ADM equations do not contain explicit time derivatives,
and thus the CFC spacetime fields are evaluated in a fully
constrained approach and at each timestep anew.

Imposing CFC in a spherically symmetric spacetime is equivalent to
solving the exact Einstein equations. For nonspherical configurations
the CFC approximation may be roughly regarded as full GR minus the
dynamical degrees of freedom of the gravitational field that
correspond to the GW content~\cite{york_71_a}. However, even
spacetimes that do not contain GWs can be not conformally flat. A
prime example are the spacetime of a Kerr black hole~\cite{garat_00_a}
or rotating fluids in equilibrium. For rapidly rotating models of
stationary neutron stars the deviation of certain metric components from
conformal flatness has been shown to reach up to $ \sim 5\% $ in
extreme cases~\cite{cook_96_a}, while the oscillation frequencies of
such models typically deviate even less from the corresponding values
obtained in full GR simulations~\cite{dimmelmeier_06_a}.

Due to its fully constrained nature, the CFC approximation permits a
straightforward and numerically more robust implementation of the
metric equations in coordinate systems containing coordinate
singularities (e.g., spherical polar coordinates in axisymmetry and
3D) compared to the Cauchy free-evolution scheme employed in
CCW. Furthermore, by definition it allows no constraint violations,
which is a significant benefit in cases where a perturbation is added
to the initial data.

In \textsc{CoCoNuT} we use Eulerian spherical coordinates
and assume axisymmetry for the core-collapse
simulations discussed here. The computational grids consist
of 250 logarithmically-spaced and centrally-condensed radial 
zones with a central resolution of $ 250 \mathrm{\ m} $
and 45 equidistant angular zones covering $ 90^\circ $.

%%%%%%%%%%%%%%%%%%%%%%%%%%%%%%%%%%%%%%%%%%%%%%%%%%%%%%%%%%%%%%%%%%%%%%
%%%%%%%%%%%%%%%%%%%%%%%%%%%%%%%%%%%%%%%%%%%%%%%%%%%%%%%%%%%%%%%%%%%%%%

\subsection{Equations of State}
\label{section:eos}

For calculations employing polytropes in rotational equilibrium,
we utilize the hybrid polytropic--ideal-fluid EOS first introduced
by~\cite{janka_93_a}. It was discussed and used in many 
previous studies involving polytropic iron core 
models (see, e.g., \cite{zwerger_97_a, dimmelmeier_02_a,
  dimmelmeier_02_b, dimmelmeier_05_a, shibata_04_a}). At densities
below nuclear matter density $\rho_\mathrm{nuc} =
2 \times 10^{14} \mathrm{\ g\ cm}^{-3}$, we choose a polytropic 
$\Gamma = \Gamma_1 \lesssim 4/3$ from the set
$ \{1.325, 1.320, 1.310, 1.300, 1.280\} $ corresponding to labels
$ \{\mathrm{G1}, \mathrm{G2}, \mathrm{G3}, \mathrm{G4},\mathrm{G5}\} $
(see also \sref{section:polytropes}).
Above $\rho_\mathrm{nuc}$, $\Gamma$ is set to $\Gamma_2 = 2.5$ 
to mimic the stiffening of the nuclear EOS. The hybrid EOS provides
for a smooth transition between the two density regimes. A thermal
contribution owing to shock heating after core bounce is modelled
via a $\Gamma$-law ideal gas EOS with $\Gamma_\mathrm{th}$ set to 1.5.

In our more realistic model calculations we employ the tabulated 
finite-temperature nuclear EOS by Shen et al.~\cite{shen_98_a} (Shen EOS)
in the variant of Marek~et~al.~\cite{marek_05_a} which includes
baryonic, electronic, and photonic pressure components.
The Shen EOS returns the fluid pressure (and additional thermodynamic
quantities) as a function of $(\rho,T,Y_e)$, where $T$ is the
temperature and $Y_e$ is the electron number fraction per baryon
for which we additionally solve the advection equation
\begin{equation}
  \frac{1}{\sqrt{-g}}
  \left( \frac{\partial (\sqrt{\gamma} D Y_e)}{\partial t} +
  \frac{\partial \left( \sqrt{-g} D Y_e (\alpha v^i - \beta^i) \right)}
  {\partial x^i} \right) = S_{Y_e},
  \label{eq:ye_advect}
\end{equation}
where $S_{Y_e}$ is a sink term owing to electron captures as discussed
in the next \sref{section:deleptonization}.
Since the codes operate with the specific internal energy $\epsilon$,
we determine the corresponding temperature $T$ iteratively
with a Newton--Raphson scheme and the EOS table.
All interpolations are carried out in tri-linear fashion
and the table is sufficiently densely spaced to lead to
an artifical entropy increase in adiabatic collapse by not
more than $\sim 2\%$.

%%%%%%%%%%%%%%%%%%%%%%%%%%%%%%%%%%%%%%%%%%%%%%%%%%%%%%%%%%%%%%%%%%%%
%%%%%%%%%%%%%%%%%%%%%%%%%%%%%%%%%%%%%%%%%%%%%%%%%%%%%%%%%%%%%%%%%%%%

\subsection{Deleptonization and Neutrino Pressure}
\label{section:deleptonization}

Electron capture during collapse reduces $Y_e$ (i.e.\ ``deleptonizes''
the collapsing core) and 
consequently decreases the size of the homologously collapsing
inner core that roughly has a $Y_e^{2}$ dependence (see, e.g., 
\cite{liebendoerfer_04_a}). The material of the inner core
is in sonic contact and determines the dynamics and the
GW signal at core bounce and 
in the early postbounce phases. Hence, it is important
to include deleptonization during collapse.
Since multi-D GR radiation-hydrodynamics calculations
are still computationally infeasible, in the simulations using the
Shen EOS we employ
a recently proposed scheme~\cite{liebendoerfer_05_a}
in which deleptonization is parametrized
based on data from detailed 1D radiation-hydrodynamics
calculations. For this we use result obtained with the \textsc{Vertex}
code and the standard set of electron capture 
rates on free protons and protons bound in heavy nuclei
\cite{rampp_00_a}. Following~\cite{liebendoerfer_05_a},
deleptonization is turned off at core bounce (defined as the point in
simulation time when the specific entropy per baryon $s$ at the
edge of the inner core reaches $3 k_\mathrm{B}$).
After core bounce $Y_e$ is passively advected,
but the postbounce deleptonization of the PNS is
not tracked.

Neutrino pressure is included
only in the neutrino optically-thick regime 
(at $\rho \gtrsim \rho_\mathrm{trapping} = 
2 \times 10^{12} \mathrm{\ g\ cm}^{-3}$)
where we treat the neutrino pressure contribution $P_\nu$
as an ideal Fermi gas as discussed in~\cite{liebendoerfer_05_a},
and include the radiation stress via
source terms in the momentum and energy equations.
Concretely, we add in fully-coupled Method-of-Lines-fashion
$- \alpha \sqrt{\gamma} \partial P_\nu / \partial x^i $ 
and  $- \alpha \sqrt{\gamma} \, v^i \partial P_\nu / \partial x^i $
as a source term to the right-hand side of the momentum 
and energy equation, respectively.
A detailed discussion of our implementation, identical 
in both \textsc{CoCoNuT} and CCW, can be found in~\cite{ott_06_d}.

%%%%%%%%%%%%%%%%%%%%%%%%%%%%%%%%%%%%%%%%%%%%%%%%%%%%%%%%%%%%%%%%%%%%%%
%%%%%%%%%%%%%%%%%%%%%%%%%%%%%%%%%%%%%%%%%%%%%%%%%%%%%%%%%%%%%%%%%%%%%%

\subsection{Initial Models}
\label{section:models}

%%%%%%%%%%%%%%%%%%%%%%%%%%%%%%%%%%%%%%%%%%%%%%%%%%%%%%%%%%%%%%%%%%%%%%

\subsubsection{Polytropes in Rotational Equilibrium\quad}
\label{section:polytropes}

For the comparison between CFC (\textsc{CoCoNuT})
and full GR (CCW) we employ $ \Gamma = 4/3 $ (i.e.\ $n=3$) polytropes
in rotational equilibrium that are obtained with the relativistic
extension of Hachisu's self-consistent field method~\cite{komatsu_89_a}. The polytropes
are set up with the rotation law discussed in~\cite{zwerger_97_a, dimmelmeier_02_a}
and are parametrized via the differential rotation parameter
$A$ and the initial ratio $\beta_\mathrm{i}$ of rotational 
kinetic energy $T$ to gravitational binding energy $|W|$.
We employ the model names of~\cite{zwerger_97_a} that relate
$ \{\mathrm{A1}, \mathrm{A2}, \mathrm{A3}, \mathrm{A4}\} $ with
$ \{A = 50,\!000 \mathrm{\ km}, A = 1000 \mathrm{\ km},
A = 500 \mathrm{\ km}, A = 100 \mathrm{\ km}\}$ and
$ \{\mathrm{B1}, \mathrm{B2}, \mathrm{B3}, \mathrm{B4}, \mathrm{B5}\} $
with $ \{\beta_\mathrm{i} = 0.25\%, \beta_\mathrm{i} = 0.50\%,
\beta_\mathrm{i} = 0.90\%, \beta_\mathrm{i} = 1.80\%,
\beta_\mathrm{i} = 4.00\%\} $. Note that the degree of differential
rotation decreases with increasing $A$.
The model names are constructed based on the values of
$A$ and $\beta_\mathrm{i}$ employed and on the value
of the polytropic index $\Gamma_1$ used in the subnuclear regime (see
\sref{section:eos}). Hence, model A1B3G5, for example, has
$ A = 50,\!000 \mathrm{\ km} $, $ \beta_\mathrm{i} = 0.90 \% $ and is
evolved with $ \Gamma_1 = 1.280 $.

For the comparison study we choose models A1B3G3 (yielding Type~I
dynamics and waveform), A3B3G2 (Type~I/II transitional),
A2B4G1 (Type~II), and A1B3G5 (Type~III) as a representative
subset of the models considered by~\cite{dimmelmeier_02_b}.

%%%%%%%%%%%%%%%%%%%%%%%%%%%%%%%%%%%%%%%%%%%%%%%%%%%%%%%%%%%%%%%%%%%%%

\subsubsection{Presupernova Models from Stellar Evolutionary Calculations\quad}
\label{section:stellar_evolution}

All presupernova stellar models available to-date are end
products of Newtonian spherically symmetric (1D) stellar evolutionary
calculations from hydrogen burning on the main sequence to the
onset of core collapse by photo-dissociation of heavy nuclei and
electron captures (see, e.g., \cite{woosley_02_a}).
Recently, the first presupernova models that include rotation in
an approximate 1D fashion have become 
available~\cite{heger_00_a, heger_05_a}. 
Here, we employ a solar-metallicity 
$20\,M_\odot$ (at zero-age main sequence) model
of~\cite{woosley_02_a} (in the following, model s20) 
and set it into rotation according to the rotation law 
specified in~\cite{zwerger_97_a} and with the same rotation nomenclature
employed for the above polytropes (such as A2B4). In addition, 
we perform calculations with the ``rotating'' presupernova model
E20A of~\cite{heger_00_a}, which we map onto our computational
grids under the assumption of constant rotation on
cylindrical shells of constant coordinate radius.
We point out that due to their 1D nature, 
none of the considered models
are in rotational equilibrium. This should not 
limit the quality of our results, since the collapse
proceeds slowly on a timescale of more than $ 100 \mathrm{\ ms} $ and,
hence, the star has sufficient time for the adjustment to the
appropriate angular stratifications for its rate of 
rotation~\cite{moenchmeyer_91_a, zwerger_97_a}. Initial curvature data
in CCW are obtained via the Newtonian metric 
approximation~\cite{schutz_85_a} and in \textsc{CoCoNuT}
via the CFC ADM equations.

In this study, we focus on the collapse of massive
presupernova iron cores with at most moderate differential
rotation and maximum precollapse rotation rates that lead
to PNSs that are likely spinning too fast to yield cold
NS spin periods in agreement with observationally
inferred injection periods of young pulsars into 
the $P/\dot{P}$ diagram~\cite{heger_05_a, ott_06_a}.
However, they may be relevant in the collapsar-type
gamma-ray burst scenario~\cite{ott_06_a, woosley_06_a}.

In \tref{table:shen_model_summary}, we summarize the parameters of the
models including microphysics calculated in this work.

\begin{table}
  \caption{Summary of the models including microphysics.
    $ \rho_\mathrm{b} $ is the density at bounce, the maximum
    characteristic GW strain~\cite{flanagan_98_a}
    $ h_\mathrm{char,max} $ is at a distance of $ 10 \mathrm{\ kpc} $,
    and $ E_\mathrm{gw} $ is the energy emitted in GWs (see, e.g.,
    \cite{dimmelmeier_05_a}). Models s20A2B4 and E20A are evolved to
    $ \sim 90 \mathrm{\ ms} $ and $ \sim 70 \mathrm{\ ms} $ after
    bounce, respectively. The table columns with model names that have
    the subscript `axi' include only the GW emission from their
    axisymmetric phases up to $\sim 10\mbox{\,--\,}20 \mathrm{\ ms}$
    after bounce.\label{table:shen_model_summary}}
  \begin{indented}
    \item[]\begin{tabular}{@{}lrrrrrr}
      \br
      Model
      & \multicolumn{1}{c}{$ A $}
      & \multicolumn{1}{c}{$ \beta_\mathrm{i} $}
      & \multicolumn{1}{c}{$ \beta_\mathrm{b} $}
      & \multicolumn{1}{c}{$ \rho_\mathrm{b} $}
      & \multicolumn{1}{c}{$ h_\mathrm{char,max} $}
      & \multicolumn{1}{c}{$ E_\mathrm{gw} $} \\
      & \multicolumn{1}{c}{[$ 10^8 \mathrm{\,cm} $]}
      & \multicolumn{1}{c}{[\%]}
      & \multicolumn{1}{c}{[\%]}
      & \multicolumn{1}{c}{$ \displaystyle \left[
        \frac{10^{14} \mathrm{\,g}}{\mathrm{cm}^3} \right] $}
      & \multicolumn{1}{c}{[$ 10^{-21} $]}
      & \multicolumn{1}{c}{[$ 10^{-9} M_\odot c^2 $]} \\ [0.7 em]
      \mr
      $\rule{0 em}{1 em}$%
      s20A1B1 & 50.0  & 0.25 &  0.90 & 3.29 &  1.46 &  0.6 \\
      s20A1B5 & 50.0  & 4.00 & 10.52 & 2.90 &  9.68 & 26.9 \\
      s20A2B2 &  1.0  & 0.50 &  6.72 & 3.07 &  8.77 & 22.0 \\
      s20A3B3 &  0.5  & 0.90 & 16.57 & 2.33 &  4.58 & 12.4 \\
      \mr
      $\rule{0 em}{1 em}$%
      s20A2B4$_\mathrm{axi}$ &  1.0 & 1.80 & 16.33 & 2.35 &  4.28 &  9.4 \\
      s20A2B4 & & & & & 64.23 & 169.0 \\
      \mr
      $\rule{0 em}{1 em}$%
      E20A$_\mathrm{axi}$ & --- & 0.37 & 11.31 & 2.79 & 12.18 & 36.9 \\
      E20A & & & & & 24.23 & 75.4 \\
      \br
    \end{tabular}
  \end{indented}
\end{table}

%%%%%%%%%%%%%%%%%%%%%%%%%%%%%%%%%%%%%%%%%%%%%%%%%%%%%%%%%%%%%%%%%%%%%%%
%%%%%%%%%%%%%%%%%%%%%%%%%%%%%%%%%%%%%%%%%%%%%%%%%%%%%%%%%%%%%%%%%%%%%%%

\subsection{Gravitational Wave Extraction}
\label{section:wave_extraction}

We employ the Newtonian quadrupole formula in the
first-moment of momentum density formulation as discussed
in~\cite{dimmelmeier_05_a} to extract the GWs generated
by aspherical accelerated fluid motions. We point
out that although the quadrupole formula is not gauge invariant
and is only valid in the Newtonian slow-motion
limit, it yields results that agree very well in phase
and to $\sim 10 \mbox{\,--\,} 20\% $ in amplitude with more
sophisticated methods~\cite{shibata_03_a}.

%%%%%%%%%%%%%%%%%%%%%%%%%%%%%%%%%%%%%%%%%%%%%%%%%%%%%%%%%%%%%%%%%%%%%%
%%%%%%%%%%%%%%%%%%%%%%%%%%%%%%%%%%%%%%%%%%%%%%%%%%%%%%%%%%%%%%%%%%%%%%
%%%%%%%%%%%%%%%%%%%%%%%%%%%%%%%%%%%%%%%%%%%%%%%%%%%%%%%%%%%%%%%%%%%%%%

\section{Results}
\label{section:results}

%%%%%%%%%%%%%%%%%%%%%%%%%%%%%%%%%%%%%%%%%%%%%%%%%%%%%%%%%%%%%%%%%%%%%%
%%%%%%%%%%%%%%%%%%%%%%%%%%%%%%%%%%%%%%%%%%%%%%%%%%%%%%%%%%%%%%%%%%%%%%

\subsection{Simple Hybrid EOS: Comparison between CFC and Full GR}
\label{section:hybrid_comparison}

In the following, we compare the time evolution of the maximum
rest-mass density $\rho_\mathrm{max}$ and the
GW signals of a set of models obtained in CFC with \textsc{CoCoNuT} 
and in full GR with CCW. When carrying out such a comparison 
of generally coordinate-dependent quantities, it is important
to understand in what way differences in the gauge conditions
might affect the comparison. Since any spherically symmetric metric
can be expressed as conformally-related to the flat-space metric, CFC
is an exact representation of full GR in spherical symmetry (see
\sref{section:coconut}). Hence, we compare the CFC gauge with the
gauge conditions employed in CCW in a spherically symmetric
calculation and find that both slicing and spatial coordinate
conditions match very well, independent of the assumption of spherical
symmetry in the case of matter-dominated spacetimes~\cite{ott_06_d}.
It is hence safe to employ coordinate-dependent quantities for the
comparison of results from \textsc{CoCoNuT} and CCW in the stellar
core collapse scenario, provided the gauge conditions presented in
sections~\ref{section:ccw} and~\ref{section:coconut} are used.

\begin{figure}
  \centering
  \includegraphics[width = 7.5 cm]{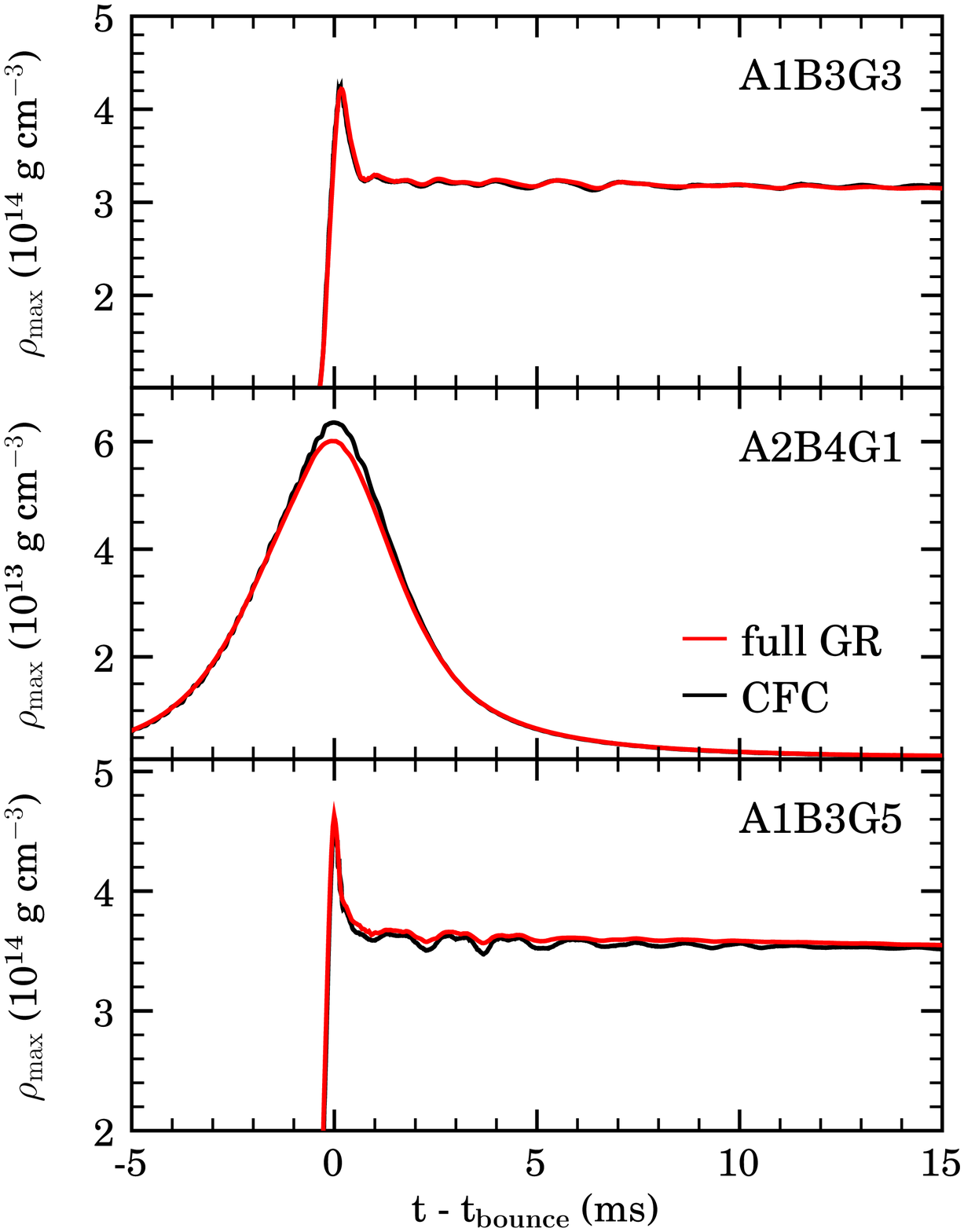}
  \hspace*{0.4 cm}
  \includegraphics[width = 7.5 cm]{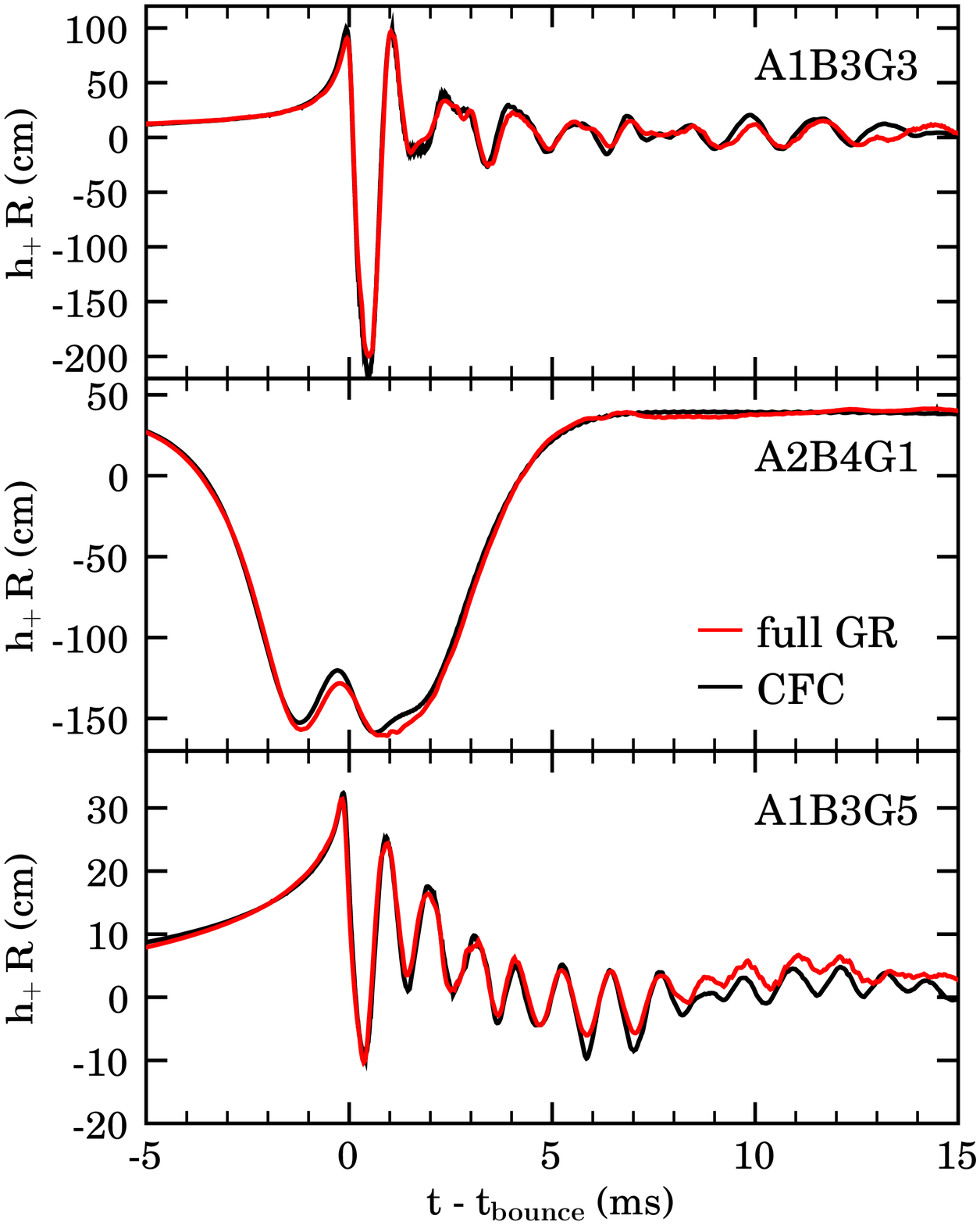}
  \caption{Comparison of the time evolution of $\rho_\mathrm{max}$ and
    of GW signals (in units of cm; rescaled by observer distance $R$)
    obtained in full GR (red curves) and CFC (black curves) for models
    A1B3G3 (Type~I), A2B4G1 (Type~II), and A1B3G5 (Type~III).}
  \label{fig:cfc_comp1}
\end{figure}

In \fref{fig:cfc_comp1}, we compare time evolutions of $\rho_\mathrm{max}$
and GW signals obtained in CFC and full GR for models
A1B3G3 (Type~I), A2B4G1 (Type~II), and A1B3G5 (Type~III). The
CFC and full GR results agree very well for all models,
modulo small differences varying from model to model 
in the degree to which small features in $\rho_\mathrm{max}$ 
and in the waveforms are resolved. These differences are 
most likely due to (i) differences in the numerical implementations,
(ii) different artificial numerical damping due to the different
choices of computational coordinates (spherical vs. Cartesian), and
(iii) slight under-resolution of the dynamics by either CCW or
\textsc{CoCoNuT}. Importantly, our results show no signs of errors in the
dynamics that could be caused by the CFC approximation of GR in 
core collapse. This finding is in agreement with previous
results~\cite{shibata_04_a, cerda_05_a}.

\begin{figure}[t]
\centering
\includegraphics[width=10.0cm]{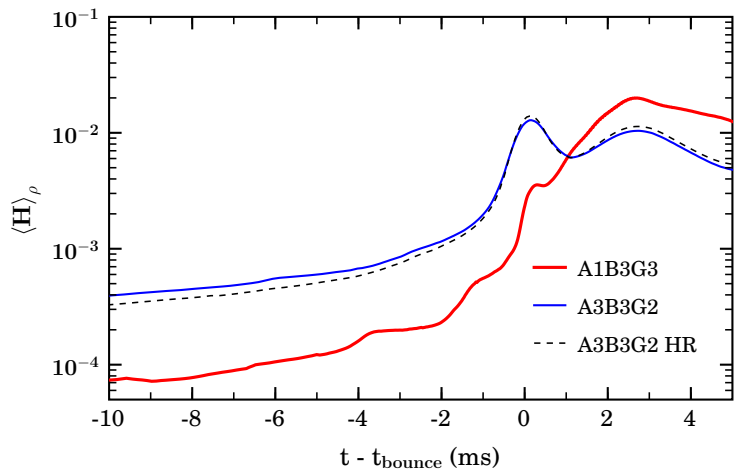}
\caption{Density-weighted volume integral of the matrix norm of the
Cotton--York tensor~(\ref{eq:cy1}, \ref{eq:cy2}) at times shortly
before and after core bounce in models A1B3G3 and A3B3G2. In
addition, we plot $\langle H \rangle_\rho$ in a calculation of
model A3B3G2 with a resolution increased overall by 20\% (labeled
HR). The numerics are reasonably well converged. Note that the
rotationally more flattened Type~I/II transitional model A3B3G2
achieves a higher prebounce value of $\langle H \rangle_\rho$, but, owing to
smaller compactness, a lower postbounce value of $\langle H \rangle_\rho$
than the Type~I model A1B3G3.}
\label{fig:cfc_comp2}
\end{figure}

In order to more quantitatively assess how far a given core-collapse
spacetime deviates from conformal flatness, we calculate
the Cotton--York tensor~\cite{york_71_a, miller_03_a, ott_06_b}
\begin{equation}
  Y^{ij} = \epsilon^{ilm} \nabla_{\!l}
  \left( R_m^j - \frac{1}{4} \delta_m^j R \right),
  \label{eq:cy1}
\end{equation}
which vanishes on conformally-flat slices. Following~\cite{miller_03_a},
we use its matrix norm normalized by the covariant derivative
of the 3-Ricci tensor,
$H = \|Y_{ij}\|/(\nabla_{\!i} R_{jk} \, \nabla^i R^{jk})$, to 
construct the rest-mass density weighted integral quantity
\begin{equation}
  \langle H \rangle_\rho = \frac{\int \mathrm{d}^3 x \, H \sqrt{\gamma} \rho W}
  {\int \mathrm{d}^3 x \, \sqrt{\gamma} \rho W}
\label{eq:cy2}
\end{equation}
as a measure of the deviation from conformal flatness. 
We perform test calculations with the Cotton--York analysis 
for models A1B3G3 and A3B3G2 and plot the time evolution of
$\langle H \rangle_\rho$ in \fref{fig:cfc_comp2}. If one
can (in a quantitative way) trust the normalization
proposed by~\cite{miller_03_a}, then the core-collapse
spacetimes studied here deviate from conformal flatness
by not more than a few percent at and shortly after core bounce (which
is in agreement with previous work~\cite{cook_96_a, cerda_05_a}. 
The situation is very likely to be different at later postbounce 
times (not considered here) when the PNS has accreted a significant 
amount of mass, cooled, and shrunk, and is thus more compact.

%%%%%%%%%%%%%%%%%%%%%%%%%%%%%%%%%%%%%%%%%%%%%%%%%%%%%%%%%%%%%%%%%%%%%%%%%%%
%%%%%%%%%%%%%%%%%%%%%%%%%%%%%%%%%%%%%%%%%%%%%%%%%%%%%%%%%%%%%%%%%%%%%%%%%%%

\subsection{Calculations with Presupernova Models and Microphysics: \\
  Axisymmetric Collapse Dynamics and Waveforms}
\label{section:shen_comparison_axi}

In this section, we present new results
for the GW signature of the axisymmetric
rotating collapse, core bounce, and early
postbounce epoch of core-collapse supernovae.
These results are obtained with CCW,
the Shen EOS, deleptonization during collapse,
and neutrino pressure effects, as delineated in
\sref{section:methods}. For all models,  
counterpart calculations are carried out
with \textsc{CoCoNuT} in axisymmetry for verification. 
The results compare as well as (or better than)
those discussed for the models with simple hybrid EOS in the previous
\sref{section:hybrid_comparison} (see also~\cite{ott_06_c}).

A first and important result of our study is that all models 
considered and listed in \tref{table:shen_model_summary} 
remain essentially axisymmetric during collapse, bounce, and the
immediate postbounce phase (most models are evolved
to $\sim 10\mbox{\,--\,}20\mathrm{\ ms}$ after bounce). The
highest $\beta$ reached at core bounce is $\sim$16.6\%.
This is much below the threshold value of $\beta \sim 27\%$
for the high-$T/|W|$ dynamical rotational bar-mode instability
and only slightly above the threshold for secular (viscosity
and/or GW-reaction driven) rotational instability that would
develop on much longer timescales than considered here.
Furthermore, and in agreement with previous 
results~\cite{tohline_84_a, ott_04_a, ott_06_a}, our
calculations indicate a natural centrifugal barrier that
limits the maximum value of $\beta$ that can be reached
during core collapse when a microphysical EOS and
presupernova models from stellar evolution calculations
are employed.

\begin{figure}
  \centering
  \includegraphics[width = 11.5 cm]{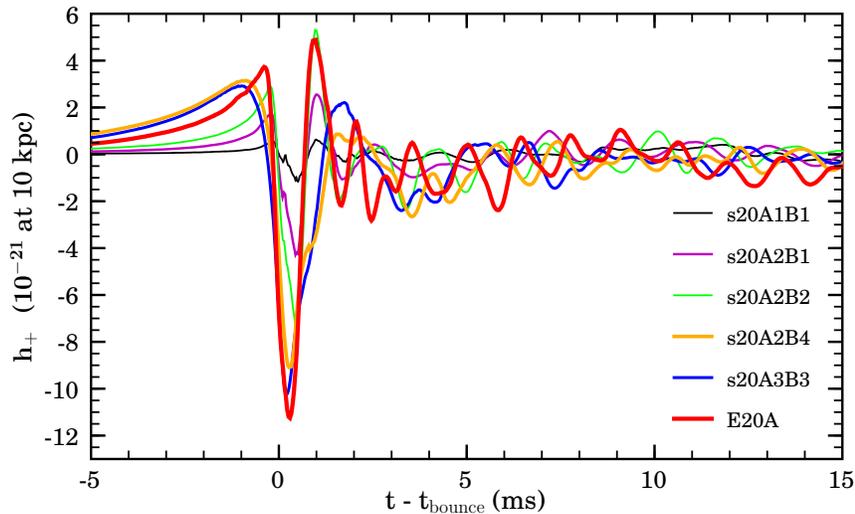}
  \caption{GW strain $h_+$ along the equator for all model calculations
    employing presupernova models from stellar evolutionary studies,
    the Shen EOS, deleptonization during collapse, and neutrino
    pressure in the neutrino optically-thick regime. Note the generic
    shape of the waveforms that is largely independent of precollapse
    rotation rate and degree of differential rotation in the parameter
    space chosen here.}
  \label{fig:waveforms}
\end{figure}

In \fref{fig:waveforms}, we present gravitational waveforms 
of models with varying initial degree of differential rotation 
$A$ and rotation rate $\beta_\mathrm{i}$. This figure 
demonstrates that largely independent of the initial 
rotational configuration in the parameter space considered 
here,  the GW signal of core bounce in rotating
collapse has a \emph{generic shape}: a slow signal increase 
in the prebounce phase, a large negative amplitude at core bounce 
when the motion of the quasi-homologously collapsing inner core 
is reversed, followed by a ring-down. This is a clear Type~I signature 
corresponding to a baryonic pressure-dominated bounce. All our
microphysical models undergo core bounce dominated by the stiffening of the EOS at nuclear
density, and most of them radiate the largest fraction of their GW
energy in a relatively narrow band (with
$\delta f \sim 50 \mathrm{\ Hz}$) centered in the range between about
$ 650 $ and $ 800 \mathrm{\ Hz}$. With increasing influence of
centrifugal effects, the peaks of the GW energy spectra get shifted to
lower frequencies, reaching $ \sim 300 \mathrm{\ Hz} $ for the
axisymmetric emission in model s20A2B4.

The above is in stark contrast to the studies using the hybrid 
EOS~\cite{zwerger_97_a, dimmelmeier_02_b, shibata_04_a, shibata_05_a} (see also
\sref{section:hybrid_comparison}), where several initial models with
rotation rates in the range investigated
here develop sufficient centrifugal support during contraction to stop
the collapse at subnuclear densities, resulting in several consecutive
centrifugal bounces separated by phases of coherent re-expansion of
the inner core. While in GR such models exhibiting a multiple
centrifugal bounce and the corresponding Type~II GW signals are
only rarer compared to Newtonian gravity~\cite{dimmelmeier_02_b} if a hybrid
EOS is used (which is identical to a polytrope before bounce), we do
not observe \emph{any} such model in our microphysical models. An
evident example is model s20A2B4: In previous studies 
without microphysics, the corresponding model with 
identical initial rotation parameters and a subnuclear $\Gamma_1$ 
close to 4/3 like A2B4G1 with
$ \Gamma_1 = 1.325 $ (or A2B4G2 with $ \Gamma_1 = 1.320 $ not
presented here) showed clear Type~II behavior in both Newtonian and GR
calculations~\cite{zwerger_97_a, dimmelmeier_02_b}.

The suppression of the multiple centrifugal bounce scenario is due to
two physical effects: On the one hand, GR results in a
stronger gravitational pull as compared to a purely Newtonian
treatment, thus forming a smaller, more compact PNS with higher
maximum density. This \emph{effectively stronger gravity} in GR
severely limits the region in rotational parameter space which permits
multiple centrifugal bounces~\cite{dimmelmeier_02_b}. Additionally,
and even more importantly, in contrast to the simple hybrid EOS, in
our case the mass and dynamics of the inner core (which is most
important for the GW emission) is not merely determined by the
adiabatic index $ \Gamma = d \ln P / d \ln \rho $ (at constant
entropy) of the EOS, but also by deleptonization during collapse. This
leads to \emph{considerably smaller inner cores} with less angular
momentum and weaker pressure support~\cite{hix_03_a}, which again
suppresses multiple centrifugal bounces in a very efficacious
way. However, even when including the effects of deleptonization, the
mass of the inner core is still substantially larger than when a
hybrid EOS with $ \Gamma_1 \lesssim 1.300 $ at subnuclear densities is
used. This explains why we do not observe any Type~III waveforms in
the microphysical models, as these require an extremely small mass of
the inner core~\cite{zwerger_97_a, dimmelmeier_02_b}.

An instructive example for the impact of deleptonization and GR
on the collapse dynamics is the Type~II model~D from 
the Newtonian study of M\"onchmeyer \& M\"uller~\cite{moenchmeyer_91_a}, 
where a microphysical finite-temperature EOS and an
approximate deleptonization scheme were employed (including only
electron capture on protons and neglecting captures on heavy nuclei
which leads to larger $Y_e$ and inner core masses 
at core bounce than observed in our models). When abandoning the
restrictions of the deleptonization scheme or gravity in that work by
using a more modern prescription for electron
capture~\cite{rampp_00_a} or including GR effects (and certainly by
combining both), we now find that their model~D exhibits unambigous
pressure-dominated collapse dynamics with a Type~I GW signal. A
detailed analysis of the interplay and quantitative influence of the
above two effects responsible for the elimination of multiple
centrifugal bounces in the rotating stellar core-collapse scenario is
discussed in~\cite{ott_06_d} and will be presented in a future
publication~\cite{dimmelmeier_07_a}.

%%%%%%%%%%%%%%%%%%%%%%%%%%%%%%%%%%%%%%%%%%%%%%%%%%%%%%%%%%%%%%%%%%%%%
%%%%%%%%%%%%%%%%%%%%%%%%%%%%%%%%%%%%%%%%%%%%%%%%%%%%%%%%%%%%%%%%%%%%%

\subsection{Calculations with Presupernova Models and Microphysics: \\
  Nonaxisymmetric Dynamics}
\label{section:shen_comparison_nonaxi}

In recent studies dynamical rotational 
instabilities of $m=2$ and/or $m=1$ character where found
in equilibrium polytropic stellar models 
(in Newtonian gravity and GR; see, e.g., \cite{saijo_06_a, ou_06_a,
shibata_04_a, zink_06_a}, 
and references therein) and in simplified postbounce
PNS models~\cite{ott_05_a} at low $\beta$ in the range of
$\sim 1\mbox{\,--\,}15\%$. In contrast to the classical high-$T/|W|$
instability, this new kind of dynamical instability appears to be related
to resonant amplification of azimuthal fluid modes at
corotation points, where the pattern speed of the mode
$\sigma_\mathrm{p} = \sigma/m$ (where $\sigma$ is the mode frequency),
coincides with the local angular velocity~\cite{watts_05_a, ou_06_a,
  saijo_06_a, ott_05_a}.

\begin{figure}
  \centering
  \includegraphics[width = 9.0 cm]{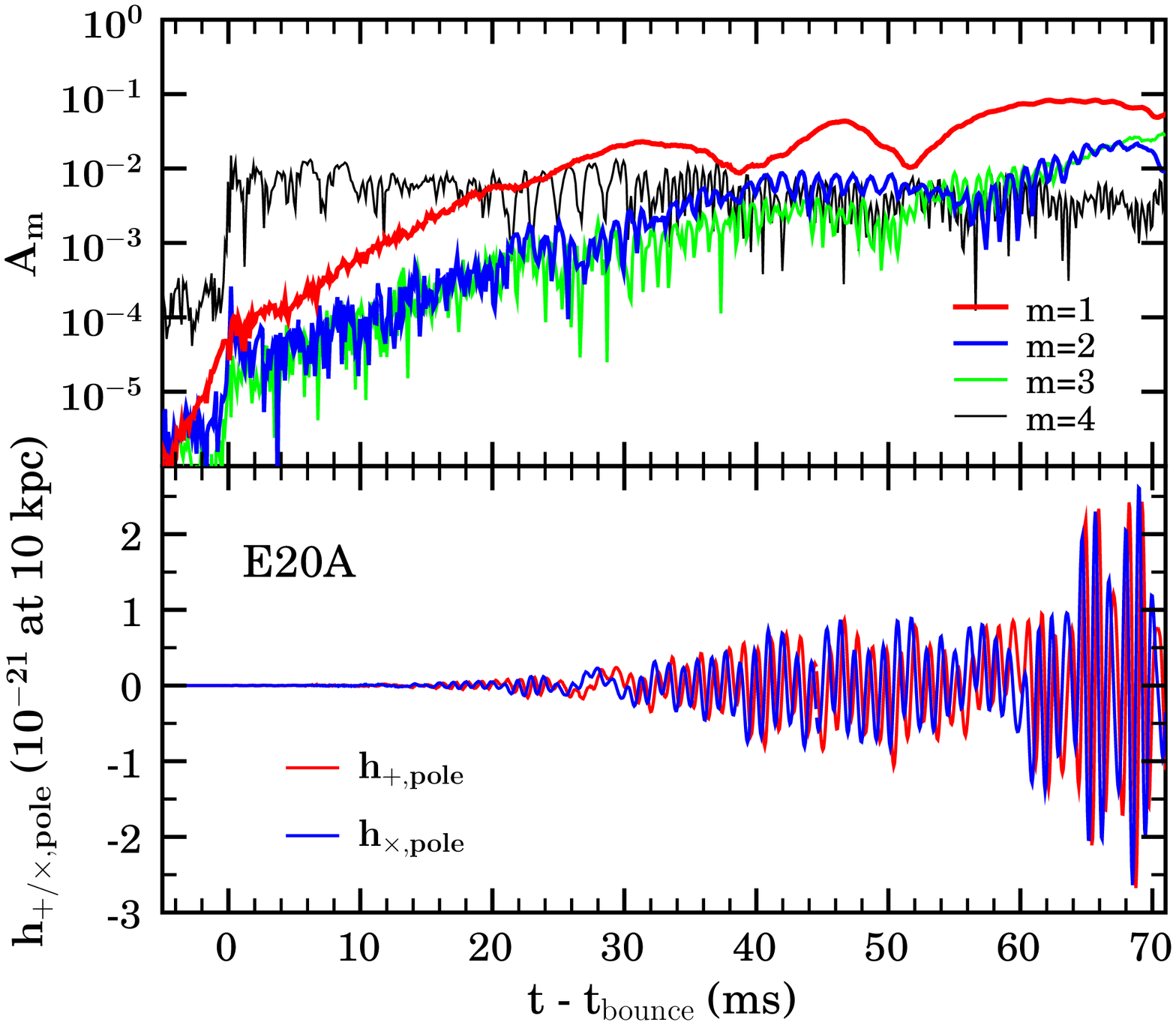} \\ [1em]
  \includegraphics[width = 9.0 cm]{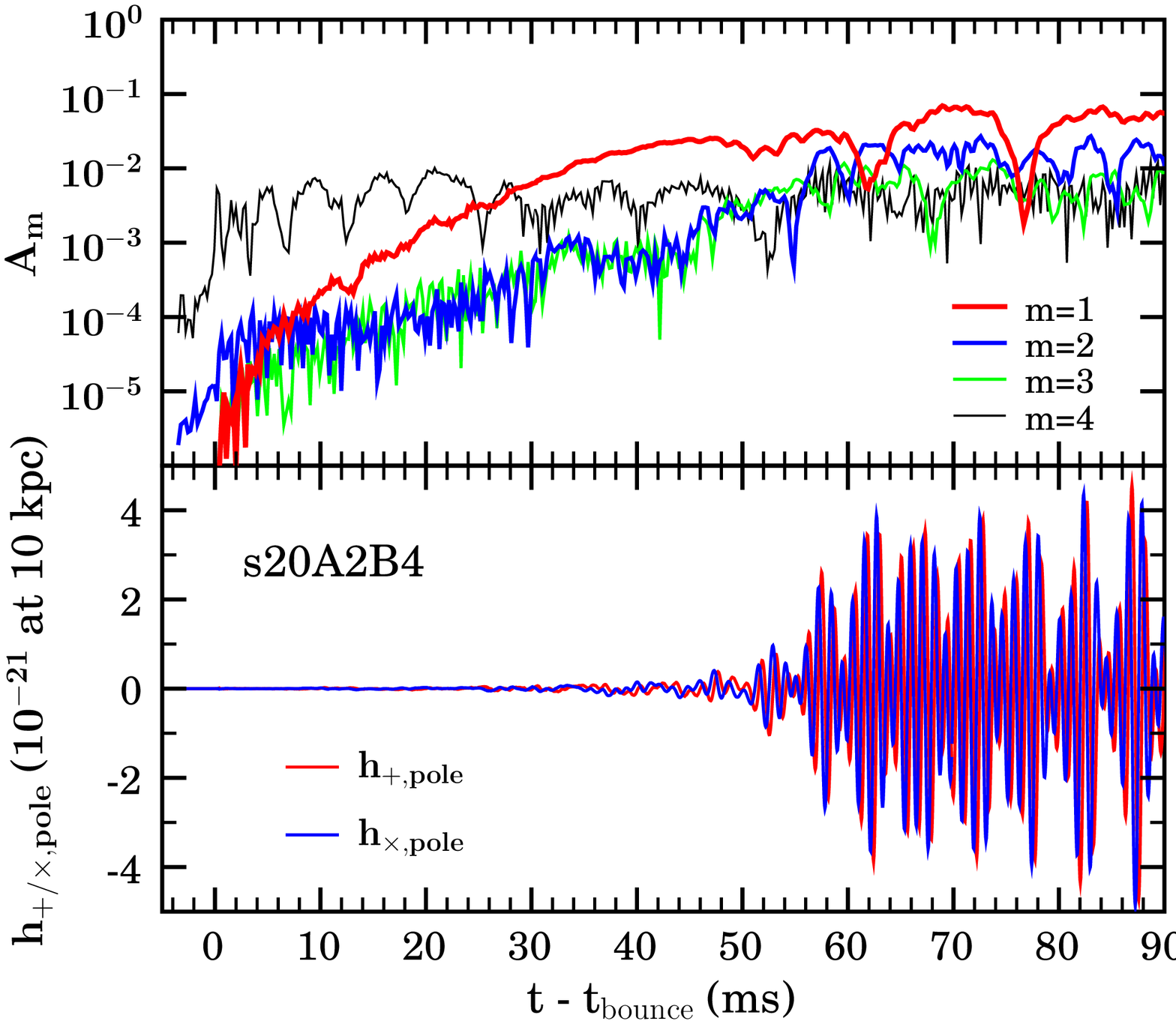}
  \caption{Normalized mode amplitudes $A_m$ in the equatorial plane
    extracted at a radius of $ 15 \mathrm{\ km} $ at postbounce times
    (upper half of the panels), and GW strains $h_+$ and $h_\times$
    along the poles (lower half of the panels) at postbounce times in
    models E20A (upper panel) and s20A2B4 (lower panel). The times are
    given relative to the time of core bounce in each model.}
  \label{fig:shen_nonaxi}
\end{figure}

In the light of the previous results 
on low-$T/|W|$ corotation instabilities, we continue
the postbounce evolution of models E20A and s20A2B4
in order to investigate the development of nonaxisymmetric
structure in the PNS and postshock regions of  
our physically more realistic models. Models E20A
and s20A2B4 are both relatively quickly spinning and bracket 
a range of postbounce $\beta$s of $\sim 9\mbox{\,--\,}13\%$.
We perform an analysis of azimuthal density modes
$ \propto \mathrm{e}^{\mathrm{i}m\varphi} $ in the equatorial plane by 
computing the complex Fourier amplitudes 
$ C_m = \frac{1}{2\pi} \int^{2\pi}_0 \rho(\varpi, \varphi, z = 0)
\, \mathrm{e}^{\mathrm{i}m\varphi} \, \mathrm{d}\varphi $ on rings of
constant coordinate radius.
In the top panels of \fref{fig:shen_nonaxi} we 
display the normalized mode amplitudes 
$ A_m = |C_m| / C_0 $ extracted in the two models at
$ 15 \mathrm{\ km} $ radius.
Without adding artificial seed perturbations to our calculations, 
discretization errors and the finite accuracy of the Newton--Raphson
scheme employed in the EOS routines trigger
$ m = \{1,2,3\} $ modes, which rise to a level of $ \sim 10^{-5} $
during the collapse phases lasting $\sim 170 \mathrm{\ ms}$ and $\sim
200 \mathrm{\ ms}$ in models s20A2B4 and E20A, respectively.

In both models, the $m=1$ mode grows fastest and with a dynamical
rate at core bounce, and surpasses the ambient Cartesian $m=4$ 
grid mode at $\sim 20\mbox{\,--\,}30 \mathrm{\ ms}$ after bounce, reaching
maximum normalized amplitudes of up to $10^{-1}$. 
$m=\{2,3\}$ modes grow as well, but at a lower rate, and 
reach smaller amplitudes than the $m=1$ mode. Note that both 
models remain dynamically stable to the $m=4$ grid mode.

\begin{figure}
  \centering
  \includegraphics[width = 12.5 cm]{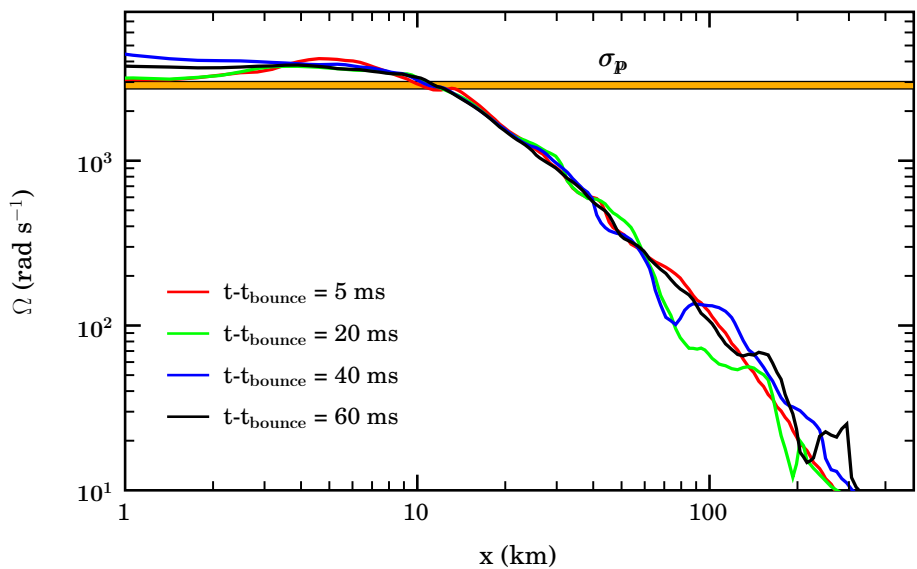}
  \caption{Angular velocity profiles along the positive $x$-axis in
    the equatorial plane in model E20A at various postbounce
    times. The pattern speed of the $m=1$ mode is marked with a
    bar. It is in corotation with the PNS at
    $ \sim 10\mbox{\,--\,}15 \mathrm{\ km} $. Note that out to
    $ \sim 10 \mathrm{\ km} $ the PNS core is in approximate solid
    body rotation. This is a consequence of the quasi-homologous
    collapse of the inner iron core~\cite{ott_06_a}. Model s20A2B4
    (not shown here) exhibits qualitatively and quantitatively similar
    features and its $m=1$ mode is in corotation in the same
    equatorial radial interval.}
  \label{fig:E20A_corot}
\end{figure}

In the lower half of the two panels of \fref{fig:shen_nonaxi} we plot
the GW strains $h_+$ and $h_\times$ as seen by an observer
located on the polar axis at $10 \mathrm{\ kpc}$ distance.
The rotational symmetry prevailing at bounce and early postbounce
times is apparent from the fact that $h_+$ as well as $h_\times$
at the pole are essentially zero until the models develop considerable
nonaxisymmetry at $\sim 30\mathrm{\ ms}$ (in E20A) and $\sim 50
\mathrm{\ ms}$ (in s20A2B4) after bounce. The GW emission along the 
polar axis is entirely due to the quadrupole components
of the nonaxisymmetric dynamics and, hence, the time at which the GW
signals become strong is correlated with the evolution of the $m=2$
mode amplitude. This fact is most clearly portrayed by model
s20A2B4 whose GW emission sets in at the 
time when the $m=2$ mode amplitude crosses that of the background
$m=4$ mode. The GW emission is in a very narrow frequency band (with
$ \delta f \sim 30 \mathrm{\ Hz} $). The energy spectra peak at
$ \sim 928 \mathrm{\ Hz} $ for model E20A and at
$ \sim 918 \mathrm{\ Hz} $ for model s20A2B4. In remarkable agreement
with expectations for a simple spinning-bar model, $h_+$ and
$h_\times$ in both models oscillate at the same frequency and are
phase-shifted by a quarter cycle.

By analyzing the nonaxisymmetric mode structure in both models
we find (i) that the pattern speeds of the $m=\{2,3\}$ modes agree
with that of the $m=1$ mode, indicating non-linear mode coupling
and (ii) that the GW emission occurs at a frequency
corresponding to twice the 
pattern speed of the $m=1$ mode. In 
\fref{fig:E20A_corot}, we compare the $m=1$ pattern speed 
with equatorial angular velocity profiles in model E20A. The
$m=1$ mode is in corotation with the fluid 
throughout the postbounce phase at the edge of the 
PNS core at $\sim 10\mbox{\,--\,}15 \mathrm{\ km}$, where the shear energy
stored in differential rotation is large. This is evidence
for the corotation low-$T/|W|$ nature of the rotational
instability observed here~\cite{saijo_06_a, ou_06_a, ott_05_a, watts_05_a}.

%%%%%%%%%%%%%%%%%%%%%%%%%%%%%%%%%%%%%%%%%%%%%%%%%%%%%%%%%%%%%%%%%%%%%%%%
%%%%%%%%%%%%%%%%%%%%%%%%%%%%%%%%%%%%%%%%%%%%%%%%%%%%%%%%%%%%%%%%%%%%%%%%
%%%%%%%%%%%%%%%%%%%%%%%%%%%%%%%%%%%%%%%%%%%%%%%%%%%%%%%%%%%%%%%%%%%%%%%%

\section{Summary and Conclusions}
\label{section:summary}

The study presented in this article relies
on $3+1$ full GR and $2+1$ CFC-GR calculations of the collapse of
rapidly rotating  stellar iron cores to PNSs. 
The $3+1$ calculations
are the first of their kind that are carried out in fully 
self-consistent fashion on mesh-refined Cartesian grids.
Our calculations are the first ever multi-D
calculations of core collapse in GR that include a
finite-temperature nuclear EOS, presupernova models
from stellar evolutionary studies, deleptonization during
collapse, and neutrino pressure effects.

By comparing results from fully GR and CFC collapse calculations we
observe no significant deviations that could be attributed to
systematic deficiencies of the CFC approximation in the stellar iron
core collapse context. By means of the Cotton--York tensor (which
vanishes on conformally-flat spacelike slices) we find that the
prebounce and early postbounce spacetimes do not deviate from
conformal flatness by more than a few percent. The CFC approximation
employed in \textsc{CoCoNuT} is, hence, an excellent choice for the
modeling of rotating core collapse. In addition, we point out that the
capability of \textsc{CoCoNuT} to perform the evolution in spherical
coordinates combined with its high computational efficiency in
axisymmetry makes it particularly well suited for studying
matter-dominated spacetimes in rotational symmetry.

Our results show that the GW signature of the
collapse, core bounce, and early postbounce phases
of the core-collapse supernova evolution is much more generic
than previously thought. We find that the dynamics
of core bounce is dominated by gravity and 
microphysics, reducing the relevance
of centrifugal support for the wide range of
initial rotation rates and degrees of differential
rotation considered here. Importantly, for our model
set we do not observe rotationally-induced multiple
core bounces as proposed by previous studies that did not
include a microphysical finite-temperature nuclear EOS
and a deleptonization treatment in combination with GR.

\begin{figure}
  \centering
  \includegraphics[width = 12.5 cm]{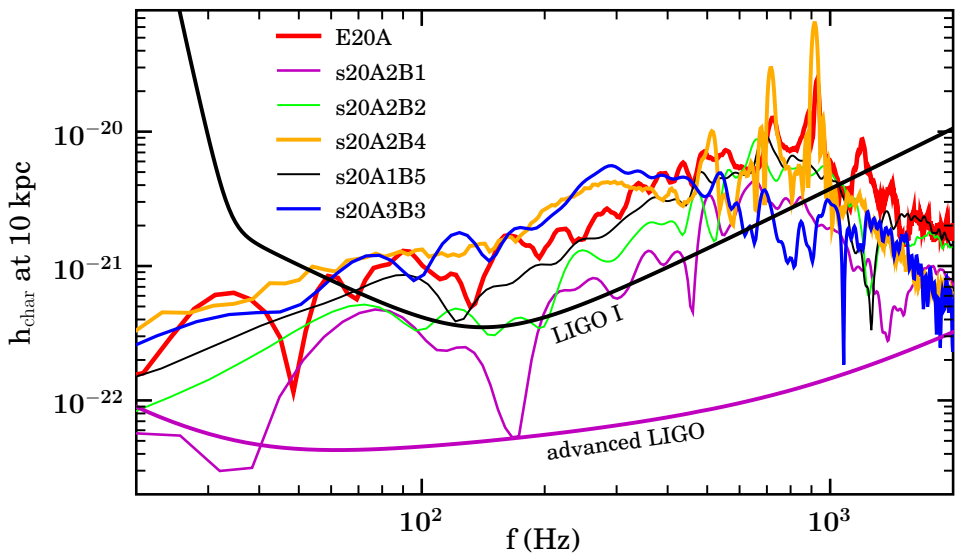}
  \caption{Frequency spectra of the characteristic GW strain
    $ h_\mathrm{char} $ of all models with microphysics (scaled to
    $ 10 \mathrm{\ kpc} $ distance) and the initial and advanced LIGO
    (optimal) rms noise curves~\cite{shoemaker_06_a}.}
  \label{fig:ligospect}
\end{figure}

All models stay axisymmetric throughout collapse, bounce, and the very
early postbounce phase, and none of them reach the limit in $ T / |W| $
for the classical dynamical MacLaurin-type rotational instability
(see, e.g., \cite{baiotti_07} for a recent study and references
therein). However, models s20A2B4 and E20A, which we evolve to later
postbounce times, exhibit the dynamical growth of a nonaxisymmetric
low-$T/|W|$ corotation-type $m=1$ instability~\cite{saijo_06_a,
ou_06_a, ott_05_a, watts_05_a}. We also observe $m=\{2,3\}$ daughter
modes that apparently grow to significant amplitudes 
via non-linear mode coupling. Strong GWs are emitted
by the quadrupole components of the nonaxisymmetric dynamics. While we
carried out long-term postbounce evolutions only for models
s20A2B4 and E20A, we point out that other models with
similar postbounce PNS structures and rotational configurations
are likely to experience the same kind of instability.
Our results, which remove the limitations of
previous studies, demonstrate that the development 
of nonaxisymmetric structures is neither limited to 
Newtonian gravity, simple matter models, 
equilibrium configurations, nor high values of $\beta$ 
above the classical instability thresholds, but may rather 
be a phenomenon occurring generically in differentially 
rotating compact stars.

For an assessment of the detectability of the GW emissions
from our models by initial and advanced LIGO detectors, 
we consider the characteristic GW strain spectra
$h_\mathrm{char} = R^{-1} \sqrt{2\pi^{-2} G c^{-3} dE_\mathrm{GW}/df}$
\cite{flanagan_98_a} and compare them
to the optimal LIGO rms noise curves~\cite{shoemaker_06_a}
in \fref{fig:ligospect}. Considering
only the axisymmetric GW burst from core bounce, $h_\mathrm{char}$
reaches values up to $\sim 10^{-20}$ and has its maximum 
between 300 and 800~Hz. The GW emission from the
nonaxisymmetric dynamics tracked in models E20A
and s20A2B4 occurs at lower GW amplitudes than that from
core bounce, but over a prolonged period of many tens 
of milliseconds and in almost monotone fashion centered around 
$\sim 920\mbox{\,--\,}930 \mathrm{\ Hz}$. In total emitted energy 
and in $h_\mathrm{char}$
the GW emission from the nonaxisymmetric instability dwarfs
that associated with the axisymmetric core bounce
and greatly enhances the GW detectability of these models.
Importantly, we point out that at the end of our 
simulations the nonaxisymmetric dynamics in models E20A
and s20A2B4 show no sign of decay. They could potentially 
continue for hundreds of milliseconds until the supernova explosion 
puts an end to the infusion of high angular momentum material
through the stalled shock and sufficient angular momentum is
redistributed by the instability to break corotation.

Based on \fref{fig:ligospect} we conclude that the GW signal
from axisymmetric core bounce of all models investigated here
is likely to be detectable by current and future LIGO-class
detectors from anywhere in the Milky Way. Models that 
become nonaxisymmetrically unstable may be detectable out to
much larger distances if the instability persists for a 
sufficiently long time.

We point out that owing to the nature of the approximate
deleptonization treatment employed in this study we are
unable to capture postbounce neutrino effects and cannot
track the sudden drop in electron fraction inside the PNS
core associated with the neutrino burst at shock
breakout a few milliseconds after bounce. Hence, our
treatment of the late-time postbounce evolution is
of limited quality, but will be improved in future
work.

\ack
We wish to thank N.~Stergioulas for permission to use his initial data
solver and M.~Miller for providing a copy of his Cotton--York routine
which was a helpful guide in code development. We furthermore
gratefully acknowledge helpful and stimulating discussions with
A.~Burrows, L.~Dessart, C.~Meakin, E.~M\"uller, S.~Ou, L.~Rezzolla,
D.~Shoemaker, E.~Seidel, N.~Stergioulas, M.~Tiglio, J.~Tohline, and
S.~Yoshida. This research was partially supported by the DFG
(SFB/Transregio 7 and SFB 375), by the Joint Institute for Nuclear
Astrophysics sub-award no.~61-5292UA of NFS award no.~86-6004791, 
by the Center for Computation \& Technology at Louisiana State
University, and by the European Network of Theoretical Astroparticle
Physics ENTApP ILIAS/N6 under contract number RII3-CT-2004-506222.
The calculations were performed on the Peyote cluster at the
Albert-Einstein-Institut, on the Supermike cluster at Louisiana State
University, and on the Teragrid machines located at NCSA (computer
time grant no.~AST05022N).

%%%%%%%%%%%%%%%%%%%%%%%%%%%%%%%%%%%%%%%%%%%%%%%%%%%%%%%%%%%%%%%%%%%%%%%%%%%%
%%%%%%%%%%%%%%%%%%%%%%%%%%%%%%%%%%%%%%%%%%%%%%%%%%%%%%%%%%%%%%%%%%%%%%%%%%%%
%%%%%%%%%%%%%%%%%%%%%%%%%%%%%%%%%%%%%%%%%%%%%%%%%%%%%%%%%%%%%%%%%%%%%%%%%%%%

\end{document}